\documentclass[review,number,sort&compress]{elsarticle}

\usepackage{tikz}
\usepackage{float}
\usepackage{graphicx}
\usepackage{gensymb}
\usepackage{amssymb}
\usepackage{color,soul}
\usepackage{url}
\usepackage{ulem}
\usepackage{textcomp}
\usepackage[figuresright]{rotating}
\usepackage{subcaption}
\usepackage{lineno}
\usepackage{subcaption}
\usepackage{hyperref}
\usepackage{cleveref}
\usepackage{libertine}
\usepackage{siunitx}
\usepackage[printwatermark]{xwatermark}
\usepackage[a4paper, total={4.73in, 8in}]{geometry}

\definecolor{amethystbg}{rgb}{0.6, 0.4, 0.8}
\definecolor{coolgreybg}{rgb}{0.55, 0.57, 0.67}
\definecolor{babypinkbg}{rgb}{0.96, 0.76, 0.76}
\definecolor{cadmiumgreenbg}{rgb}{0.0, 0.42, 0.24}
\definecolor{bluebg}{rgb}{.63,.79,.95}
\definecolor{orangebg}{rgb}{1,0.5,0}
\colorlet{lightbluebg}{bluebg!40}
\colorlet{lightorangebg}{orangebg!40}
\colorlet{lightcadmiumgreenbg}{cadmiumgreenbg!40}
\colorlet{lightbabypinkbg}{babypinkbg!40}
\colorlet{lightcoolgreybg}{coolgreybg!40}
\colorlet{lightamethystbg}{amethystbg!40}

\DeclareGraphicsExtensions{.eps,.pdf,.png,.jpg,}

\journal{Nuclear Instruments and Methods A}

\begin{document}

\newcommand{\etal}{{\it et~al.}}
\newcommand{\geant} {{{G}\texttt{\scriptsize{EANT}}4}}
\newcommand{\srim} {\texttt{SRIM}}
\newcommand{\python} {\texttt{Python}}
\newcommand{\pandas} {\texttt{pandas}}
\newcommand{\ROOT} {\texttt{ROOT}}

\DeclareRobustCommand{\hlb}[1]{{\sethlcolor{lightbluebg}\hl{#1}}}
\DeclareRobustCommand{\hlo}[1]{{\sethlcolor{lightorangebg}\hl{#1}}}
\DeclareRobustCommand{\hlg}[1]{{\sethlcolor{lightcadmiumgreenbg}\hl{#1}}}
\DeclareRobustCommand{\hlp}[1]{{\sethlcolor{lightbabypinkbg}\hl{#1}}}
\DeclareRobustCommand{\hlgr}[1]{{\sethlcolor{lightcoolgreybg}\hl{#1}}}
\DeclareRobustCommand{\hla}[1]{{\sethlcolor{lightamethystbg}\hl{#1}}}

\begin{frontmatter}


\title{Response of a Li-glass/multi-anode photomultiplier detector to focused proton and deuteron beams}


	\author[lund]{E.~Rofors}
	\author[lund]{J.~Pallon}
	\author[ess,glasgow]{R.~Al~Jebali}
	\author[glasgow]{J.R.M.~Annand}
	\author[glasgow]{L.~Boyd}
	\author[dmsc]{M.J.~Christensen}
	\author[julichZ]{U.~Clemens}
	\author[LLB]{S.~Desert}
	\author[lund]{M.~Elfman}
	\author[julich]{R.~Engels}
	\author[lund,ess]{K.G.~Fissum\corref{cor1}}
	\ead{kevin.fissum@nuclear.lu.se}
	\author[julich]{H.~Frielinghaus}
	\author[lund]{R.~Frost}
	\author[glasgow]{S.~Gardner}
	\author[ideas]{C.~Gheorghe}
	\author[ess,milan]{R.~Hall-Wilton}
	\author[julich]{S.~Jaksch}
	\author[ess]{K.~Kanaki}
	\author[julich]{G.~Kemmerling}
	\author[lund]{P.~Kristiansson}
	\author[glasgow]{K.~Livingston}
	\author[lund,ess]{V.~Maulerova}
	\author[lund]{N.~Mauritzson}
	\author[glasgow]{R.~Montgomery}
	\author[lund]{H.~Perrey}
	\author[dmsc]{T.~Richter}
	\author[lund,ess]{J.~Scherzinger\fnref{fn2}}
	\author[glasgow]{B.~Seitz}
	\author[dmsc]{M.~Shetty}

	\address[lund]{Division of Nuclear Physics, Lund University, SE-221 00 Lund, Sweden}
	\address[ess]{Detector Group, European Spallation Source ERIC, SE-221 00 Lund, Sweden}
	\address[dmsc]{Data Management and Software Centre, European Spallation Source, Ole Maaløes Vej 3, 2200 Copenhagen, Denmark}
	\address[glasgow]{SUPA School of Physics and Astronomy, University of Glasgow, Glasgow G12 8QQ, Scotland, UK}
	\address[milan]{Dipartimento di Fisica ``G. Occhialini'', Universit\`a degli Studi di Milano-Bicocca, Piazza della Scienza 3, 20126 Milano, Italy}
	\address[LLB]{LLB, CEA, CNRS, Universit\'e Paris-Saclay, CEA-Saclay 91191 Gif-sur-Yvette, France}
	\address[julich]{J\"ulich Centre for Neutron Science JCNS, Forschungszentrum J\"ulich, D-52425 J\"ulich, Germany}
	\address[julichZ]{Zentrum f\"ur Anwendungsentwicklung und Elektronik ZEA-2, Forschungszentrum J\"ulich, D-52425 J\"ulich, Germany}
	\address[ideas]{Integrated Detector Electronics AS, Gjerdrums vei 19, NO-0484 Oslo, Norway}

	\cortext[cor1]{Corresponding author. Telephone:  +46 46 222 9677; Fax:  +46 46 222 4709}
	\fntext[fn2]{present address: Thermo Fisher Scientific Messtechnik GmbH, Frauenauracher Str. 96, 91056 Erlangen, Germany}

\begin{abstract}
	The response of a position-sensitive Li-glass based scintillation detector 
	to focused beams of 2.5\,MeV protons and deuterons has been investigated. 
	The beams were scanned across the detector in 0.5\,mm horizontal and 
	vertical steps perpendicular to the beams. Scintillation light was 
	registered using an 8~$\times$~8 pixel multi-anode photomultiplier tube. 
	The signal amplitudes were recorded for each pixel on an event-by-event 
	basis. Several pixels generally registered considerable signals at each 
	beam location. The number of pixels above set thresholds were investigated, 
	with the optimization of the single-hit efficiency over the largest 
	possible area as the goal. For both beams, at a threshold of $\sim$50\% of 
	the mean of the full-deposition peak, $\sim$80\% of the events were 
	registered in a single pixel, resulting in an effective position resolution 
	of $\sim$5\,mm in X and Y.
\end{abstract}


\newpage
\end{frontmatter}
\section{Introduction}
\label{section:introduction}

Position-sensitive $^{3}$He-free~\cite{kouzes09,shea10,zeitelhack12} 
thermal-neutron detectors with high counting-rate capability are essential to 
the scientific program to be carried out at the European Spallation Source 
(ESS)~\cite{kirstein14,ess}. \underline{S}olid-state \underline{N}eutron 
\underline{D}etectors SoNDe (patent EP000003224652A1)~\cite{sonde,jaksch17cumulative,jaksch18} 
with two-dimensional position sensitivity will be employed for small-angle 
neutron-scattering 
experiments~\cite{heiderich91,ralf97,ralf98,ralf99,kemmerling01,ralf02,kemmerling04a,kemmerling04b,jaksch14,feoktystov15}. 
The modular SoNDe concept will facilitate the instrumentation of large areas
with a position reconstruction accuracy of $\sim$6\,mm for the detected neutron. 
A SoNDe ``module" consists of a thin Li-glass scintillator sheet (GS20) that is 
sensitive to thermal neutrons coupled to a 64-pixel multi-anode photomultiplier 
tube (MAPMT) used to collect the scintillation light.  Signals are read out using 
custom electronics. 

Laser light has previously been employed to study the responses of several 
different MAPMTs in detail~\cite{korpar00,rielage01,matsumoto04,lang05,abbon08,montgomery12,montgomery13,montgomery15,wang16}. 
Thermal neutrons have been used to perform first tests both on similar 
detectors~\cite{zaiwei12} and on SoNDe modules~\cite{jaksch18,roforsIFE}. 
A thermal-neutron interaction with the $^{6}$Li of the Li-glass results in an 
$\alpha$-particle (2.05\,MeV) and a triton 2.73\,MeV). Scans of a collimated 
beam of $\sim$4\,MeV $\alpha$-particles from a $^{241}$Am source have been used 
to study the position-dependent response of a SoNDe detector 
prototype~\cite{rofors19}. Here, in the absence of tritium beams, beams of 
2.5\,MeV protons and deuterons have been scanned across the face of a SoNDe 
module. The goals were to:
\begin{enumerate}
	\item complement the existing $\alpha$-particle studies on the 
		position-sensitive response of the detector for events 
		triggering up to four pixels
	\item provide data with precision of better than 1\,mm on the 
		position sensitivity of the detector, particularly for events 
		triggering only one pixel (Sec.~\ref{subsubsection:readout} 
		and Sec.~\ref{section:results})
	\item investigate the areal response of the detector to events which 
		trigger only one pixel
	\item determine the optimal threshold for such events
\end{enumerate}

\section{Apparatus}
\label{section:apparatus}

\subsection{Proton and deuteron beams}
\label{subsection:proton_and_deuteron_beams}

The Lund Ion Beam Analysis Facility~\cite{LIBAF} of the Division of Nuclear 
Physics at Lund University employs a single-ended 3\,MV (max) Pelletron 
electrostatic accelerator supplied by the National Electrostatics Corporation 
(NEC)~\cite{NEC}. This machine was used to deliver continuous beams of protons 
and deuterons with energies of 2.5\,MeV to the module under investigation.

\begin{figure}[H]
	\begin{center}
		\includegraphics[width=1.\textwidth]{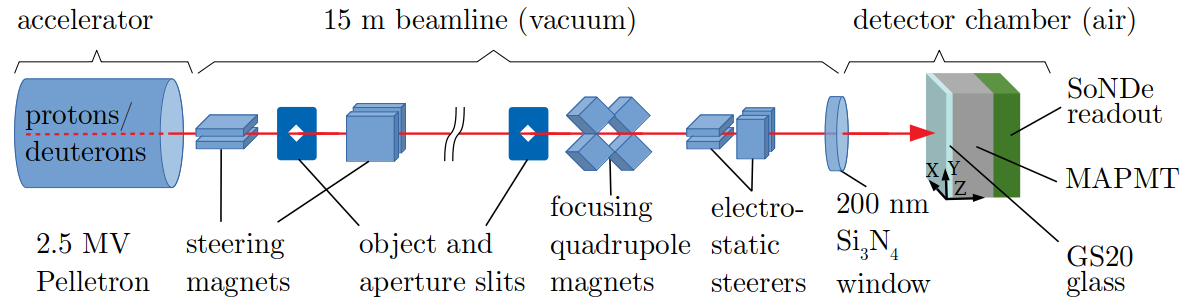}
		\caption{
			A schematic view of the experimental setup (not to 
			scale). The accelerator (left) produced continuous 
			beams of protons or deuterons with currents in the 
			$\sim$nA range. The beams were transported and 
			reduced in intensity via a beamline (middle) ending 
			in a thin vacuum window. A detector chamber (right) 
			operated at room temperature and pressure was 
			positioned downstream of this window. The detector 
			chamber contained a SoNDe module mounted on a 
			motorized platform.
			}
		\label{figure:beam}
	\end{center}
\end{figure}

Figure~\ref{figure:beam} shows the experimental setup.  A ~15 m long beamline 
between the Pelletron and the end station consisted of dipole magnets for 
energy selection and steering, object and aperture slits for adjusting the 
beam size and intensity, quadrupole magnets for focusing, and electrostatic
steerers for fine tuning of the beam position~\cite{shariff05,elfman05}.
A $\sim$200\,nm thick Si$_3$N$_4$ vacuum window~\cite{Silson} separated the 
high-vacuum beamline from the detector chamber operated at room pressure and 
temperature. The detector chamber contained a motorized XYZ translator on 
which a SoNDe module (Fig.~\ref{subfig:photograph}) was mounted. The beam 
spots at the location of the SoNDe module were estimated to be 
$\sim$\SI{100}{\micro\meter} in diameter using a fluorescent glass plate. The 
sizes of the beamspots were due largely to multiple scattering in the vacuum 
window. Beam intensity was adjusted using the aperture slits so that the 
average counting rate on the SoNDe module was 5\,kHz.

\begin{figure}[H]
	\begin{center}
		\begin{subfigure}[b]{0.49\textwidth}
			\includegraphics[width=1.\textwidth]{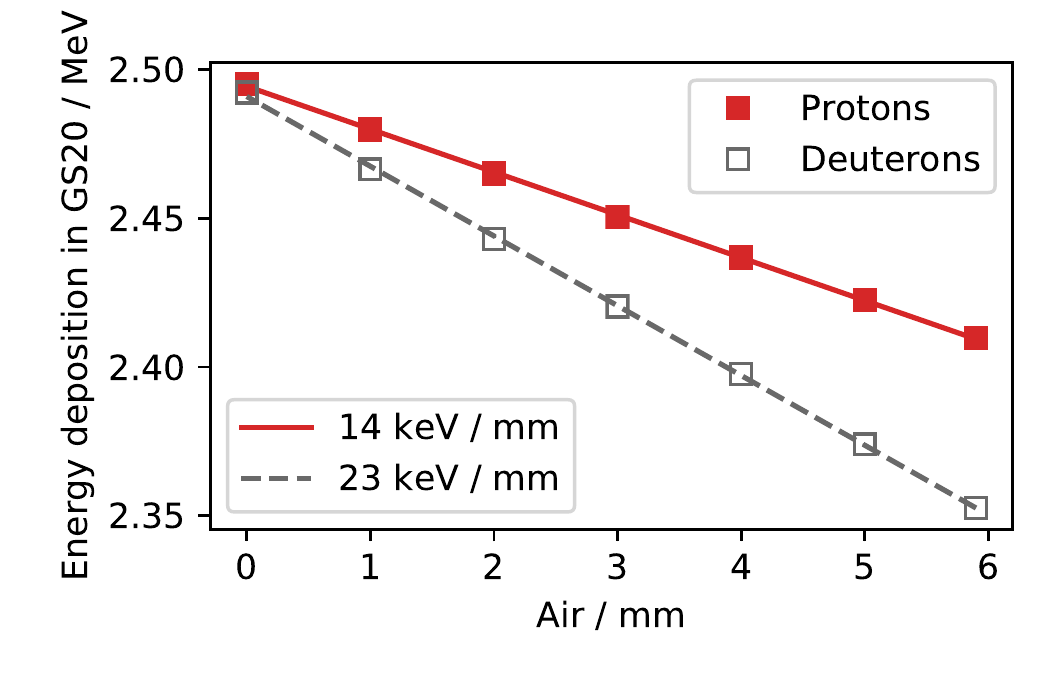}
			\caption{\srim~predictions}
			\label{subfig:SRIM}
		\end{subfigure}
		\begin{subfigure}[b]{0.49\textwidth}
			\includegraphics[width=1.\textwidth]{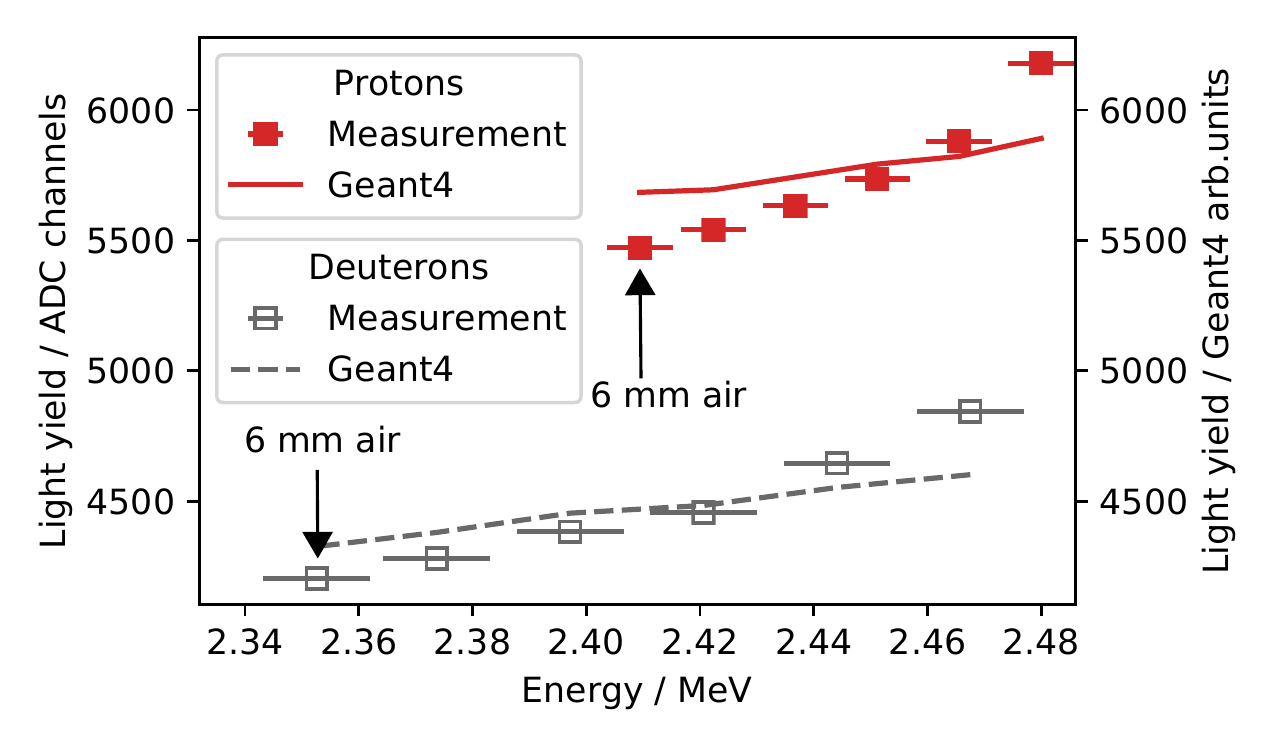}
			\caption{measurements and \geant~simulations}
			\label{subfig:Geant4_air}
		\end{subfigure}
		\caption{
			Energy deposition and scintillation-light yield.
			\ref{subfig:SRIM} shows \srim~predictions for the
			energy deposited in the scintillator by the beams 
			after passing through the vacuum window and 
			traversing increasing distances of room air before 
			striking the Li-glass scintillator.
			\ref{subfig:Geant4_air} shows measured and 
			\geant-simulated scintillation-light yields produced 
			by these beams which have systematically decreasing 
			energies. Horizontal error bars are due to the 
			uncertainty in the distance of air traversed while 
			the vertical error bars are smaller than the symbols.
			}
		\label{figure:beam_details}
        \end{center}
\end{figure}

The amplitude of proton and deuteron signals was measured as the thickness 
of the traversed air gap between the vacuum window and the GS20 was increased 
in 1\,mm steps up to 6\,mm (Fig.~\ref{figure:beam_details}). 
Figure~\ref{subfig:SRIM} shows \srim~\cite{ziegler04,SRIM} calculations of 
the proton and deuteron energy loss in the vacuum window and air. These 
predict that a 2.5\,MeV proton loses $\sim$6\,keV in the window and 
14\,keV/mm in air, while for deuterons the equivalent numbers are 
$\sim$10\,keV and 23\,keV/mm. Figure~\ref{subfig:Geant4_air} shows the 
measured scintillation-light yield from the GS20 as the air gap is varied, 
along with a \geant~simulation~\cite{annand20}. The simulation, in addition 
to energy deposition, models scintillation emission and transport. It fits the 
data best when the Birks constant~\cite{birks51,birks64} for GS20 is set to 
0.021\,mm/MeV. The light yield predicted by the simulation was normalized to 
the measured data so the deviation between measurements and simulations was 
minimized (at most 5\%). The deviation could stem from a combination of 
uncertainty in the measured air gap ($\pm$0.2\,mm) and effects not yet covered 
in the simulation. The correlation between the data and simulations confirms 
that protons of all energies produce more scintillation light than deuterons 
of the same energy. This is because for a given energy, the specific ionization 
density of deuterons is higher than that of protons, resulting in a higher 
level of saturation of the local scintillation-production mechanisms.

\subsection{SoNDe module}
\label{subsection:detector}

As described in the following sections, the core components of a SoNDe module 
(Fig.~\ref{figure:SoNDeModule}) are:
\begin{enumerate}
	\item a thin, lithium-silicate, scintillating-glass sheet
	\item a MAPMT
	\item purpose-built SoNDe readout electronics
\end{enumerate}

\subsubsection{Li-glass scintillator}
\label{subsubsection:Liglass_scintillator}

GS20~\cite{firk61,spowart76,spowart77,fairly78} is cerium-activated 
lithium-silicate glass scintillator developed for the detection of thermal
neutrons. The 50\,mm~$\times$~50\,mm~$\times$~1\,mm sheet from 
Scintacor~\cite{scintacor}, had polished front and rear surfaces and 
rough-cut 1\,mm edges. The sheet was held in place on the MAPMT window using 
tape along the thin edges. Consistent with the planned configuration at ESS,
no optical coupling medium was employed between the GS20 and MAPMT and no 
optical reflector was placed over the front face of the GS20. The density of 
$^{6}$Li in GS20 (assumed to be uniform) is 
1.58~$\times$~$10^{22}$~atoms/cm$^{3}$. At thermal energies (25\,meV), the 
n + $^{6}$Li $\rightarrow$ $^{3}$H + $\alpha$ capture cross section is 940~b, 
resulting in a detection efficiency of $\sim$75\% for a 1\,mm sheet. The 
capture process produces a 2.73\,MeV $^{3}$H (average range in GS20 of 
\SI{34.7}{\micro\meter}) and a 2.05\,MeV $\alpha$-particle (average range in 
GS20 of \SI{5.3}{\micro\meter})~\cite{jamieson15}. The 6600 photon 
scintillation-light yield~\cite{jaksch18} corresponding to a thermal-neutron 
interaction (4.78\,MeV) is quoted as 20-30\% of anthracene and the emission 
spectrum peaks at 390\,nm~\cite{vanEijk04}. For 2.5\,MeV protons and 
deuterons, the \geant~simulation predicts $\sim$125 and $\sim$100 
scintillation photons reaching the photocathode, respectively. Light 
transport from the GS20 (refractive index~1.55 at 395\,nm) through a thin air 
gap (refractive index~1) into the MAPMT borosilicate glass window (refractive 
index~1.53) is rather inefficient.

\subsubsection{Multi-anode photomultiplier tube}
\label{subsubsection:mapmt}

\begin{figure}[H]
	\begin{center}
		\begin{subfigure}[b]{0.32\textwidth}
			\includegraphics[width=1.0\textwidth]{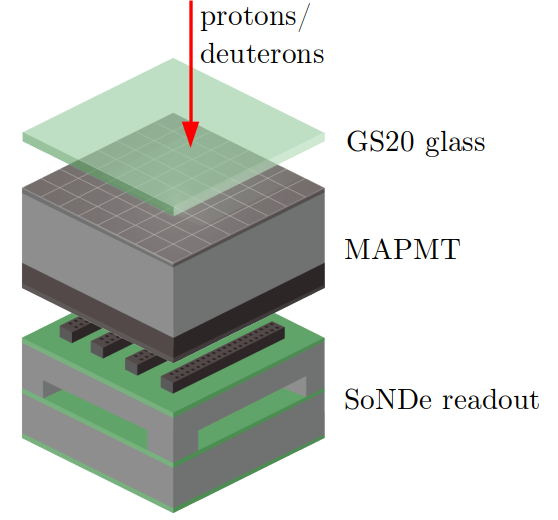}
			\vspace*{0mm}
			\caption{SoNDe module \linebreak (oblique view)}
			\label{subfig:schematic}
		\end{subfigure}
		\begin{subfigure}[b]{0.32\textwidth}
			\includegraphics[width=1\textwidth]{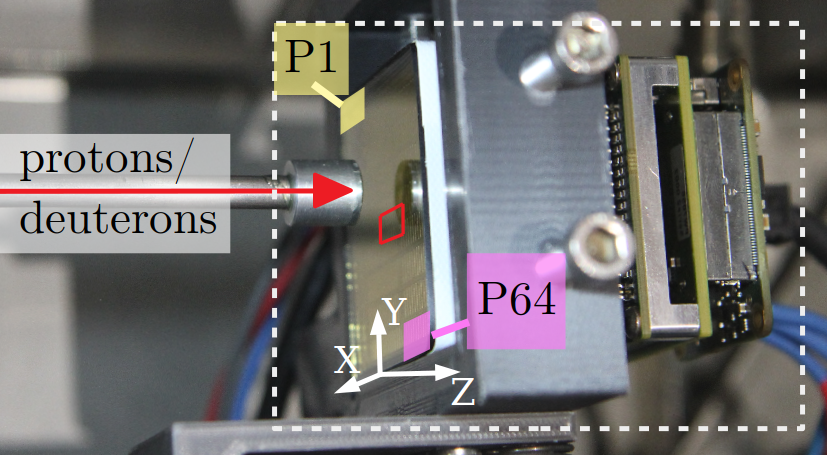}
			\vspace*{5mm}
			\caption{SoNDe module \linebreak (side view)}
			\label{subfig:photograph}
		\end{subfigure}
		\begin{subfigure}[b]{0.32\textwidth}
			\includegraphics[width=1.\textwidth]{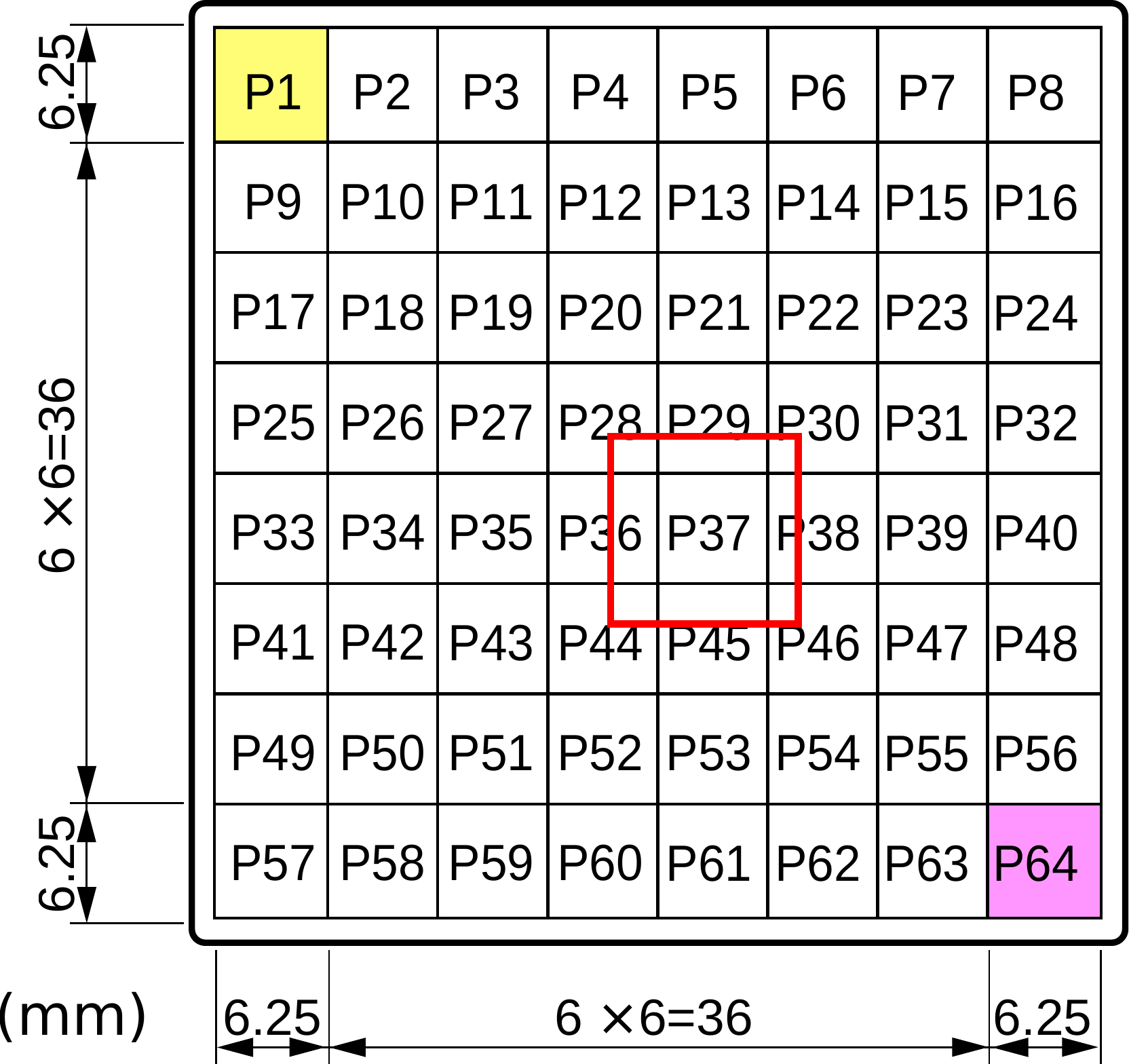}
			\caption{MAPMT pixel map \linebreak (front view)}
			\label{subfig:pixelmap}
		\end{subfigure}
		\caption{
			The SoNDe module.
			\ref{subfig:schematic}: 3D rendering of the SoNDe 
			module. From the top, scintillator, MAPMT, and
			readout electronics. Beams of protons and deuterons
			(red arrow) arrive from the top.
			\ref{subfig:photograph}: Photograph of the SoNDe 
			module (dashed white box -- from the left, 
			scintillator, MAPMT, and readout electronics) 
			mounted on the motorized platform within the 
			detector chamber downstream of the vacuum window. 
			Beams of protons and deuterons (red arrow) arrive 
			from the left.  
			\ref{subfig:pixelmap}: Numbering scheme for the 
			MAPMT pixels (front view)~\cite{hamabrochure}. For 
			orientation, Pixel~1 (P1, yellow), Pixel~64 (P64, 
			pink), and the region of systematic irradiation (red 
			box) are indicated both in panel \ref{subfig:photograph} 
			and \ref{subfig:pixelmap}. 
			(For 
			interpretation of the references to color in this 
			figure caption, the reader is referred to the web 
			version of this article.)
			}
	\label{figure:SoNDeModule}
	\end{center}
\end{figure}

Figure~\ref{subfig:photograph} shows a photograph of the SoNDe module mounted 
in the detector chamber, while Fig.~\ref{subfig:pixelmap} shows a MAPMT pixel 
map. The 8~$\times$~8 pixel Hamamatsu H12700A MAPMT chosen for the SoNDe 
module employs a borosilicate glass window. The outer dimensions of the MAPMT 
are 52\,mm~$\times$~52\,mm, while the active area of the photocathode is
48.5\,mm~$\times$~48.5\,mm. Thus 87\% of the MAPMT surface is active. Each of 
the 64 pixels has an area of $\sim$6\,mm~$\times$~$\sim$6\,mm. The peak 
quantum efficiency of the bialkali photocathode, $\sim$33\% at $\sim$380\,nm, 
is well matched to the scintillation emission spectrum from GS20, which peaks 
at $\sim$390\,nm. The Hamamatsu data sheet for the H12700A MAPMT used in this 
study gives a gain of 2.09~$\times$~10$^{6}$ and a dark current of 2.67\,nA at 
a cathode-to-anode voltage of $-$1000\,V, and a factor 1.7 (worst-case)
pixel-to-pixel gain difference.

\subsubsection{Readout electronics}
\label{subsubsection:readout}

Produced by IDEAS~\cite{ideas}, the readout electronics for the SoNDe 
module~\cite{jaksch18} consist of a front-end board and a controller board. 
The front-end board accomodates four 16-channel IDE3465 ASICs which digitize 
the MAPMT signals with 14 bit precision. The controller board houses an FPGA 
and a MiniIO port for communication via ethernet. Two modes of operation are 
``Time-of-flight" mode (TOF), envisioned for production running at ESS at 
average rates of 20~MHz/m$^{2}$ and ``All-channel Spectroscopy" mode (ACS), 
used in this work, with a rate limitation of $\sim$10\,kHz for one SoNDe 
module, equivalent to $\sim$4\,MHz/m$^{2}$. In TOF mode, when any 
pixel-amplitude threshold is exceeded, the controller board is signaled to 
identify the trigger pixel (the pixel with the largest signal), perform the 
time stamping, and then pass the resulting data to the ethernet interface. 
In ACS mode, when any pixel-amplitude threshold is exceeded, the digitized 
signals from all 64 pixels are read out. In the ACS-mode investigations, a 
low hardware threshold of 750~ADC channels was employed, which corresponds 
to 12.5\% of the mean channel of the distribution of the full energy 
deposition of 2.48~MeV protons (Fig.~\ref{figure:prot_hor}). Higher 
thresholds were applied offline, as were corrections for differing pixel 
gains.

\section{Measurement}
\label{section:measurement}

Proton and deuteron beams were used to systematically irradiate the SoNDe 
module at well-defined positions. After leaving the vacuum window, the beams 
passed through $\sim$1.0\,mm of air before striking the upstream face of the 
GS20 sheet at normal incidence. The SoNDe module was translated with its 
face perpendicular to the direction of the beams using an XYZ-coordinate 
scanner instrumented with Physik Instrumente M-111 micro translation stages 
and C-862 motor controllers~\cite{PI}. The scanning assembly was configured 
to allow for regular scans in two dimensions with a stepsize of 0.5\,mm in 
both the X and Y~directions. The assembly could also move in the Z~direction 
away from the vacuum window. The temperature ($\sim$25\,\degree C), pressure 
($\sim$101.3\,kPa), and humidity ($\sim$30\%) within the detector chamber 
that housed the scanning assembly were logged at the beginning and end of 
each scan. 

The anode signals from each of the pixels in the MAPMT were processed using
the purpose-built SoNDe electronics. The negative polarity analog pulses for 
each event with at least one pixel showing a signal above the threshold
were measured. The threshold setting corresponded to an ADC value of about 
750. The data were recorded on disk using an ESS Event Formation 
Unit~(EFU)~\cite{EFU, EFU2, EFU3} running on a Centos~7 PC connected through 
the MiniIO port to Ethernet using the UDP protocol~\cite{christensen17}. 
The EFU data-acquisition system is designed for use by ESS instruments and 
the acquisition closely resembles the mode of operation anticipated at ESS.  
Data were recorded for $\sim$2~s (10000~events) at each point on a scan, 
followed by a motor translation, so that a complete scan of 2~$\times$~2 
pixels with 0.5\,mm spacing took several hours. The data were subsequently 
analyzed using the \python-based~\cite{python} \pandas~\cite{pandas} 
analysis tools.

\section{Results}
\label{section:results} 

Previous
work~\cite{korpar00,rielage01,matsumoto04,lang05,abbon08,montgomery12,montgomery15,rofors19} 
clearly indicates that MAPMT pixel-gain maps are highly dependent upon the 
method of photon production. Thus, all of the results presented below have 
been pedestal and gain corrected with pixel-gain maps generated from the 
average of the proton- and deuteron-beam irradiations of the pixel centers.

\begin{figure}[H]
        \begin{center}
		\begin{subfigure}[b]{0.47\textwidth}
			\includegraphics[width=1.\textwidth]{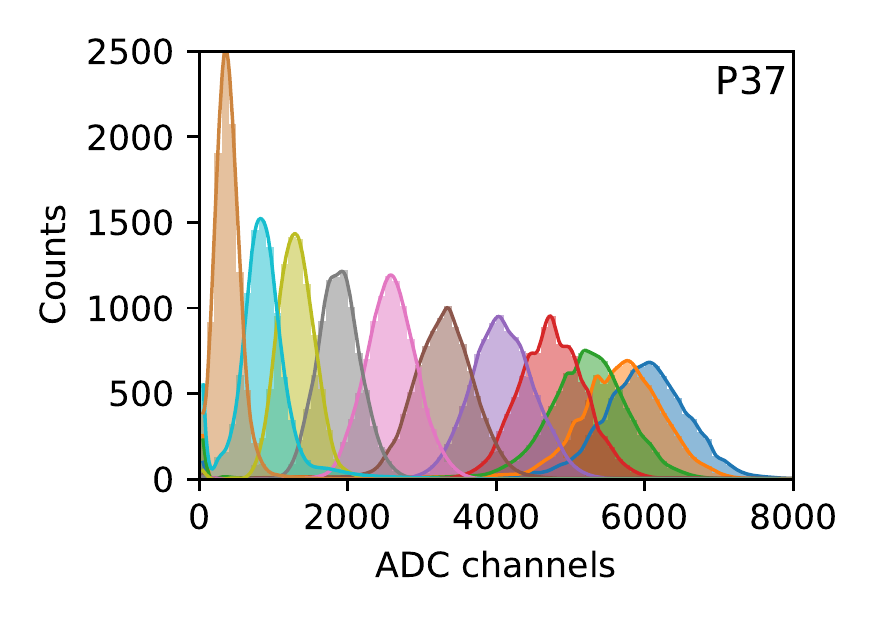}
			\caption{P37, collected charge}
			\label{subfig:prot_hor_P37}
		\end{subfigure}
		\begin{subfigure}[b]{0.47\textwidth}
			\includegraphics[width=1.\textwidth]{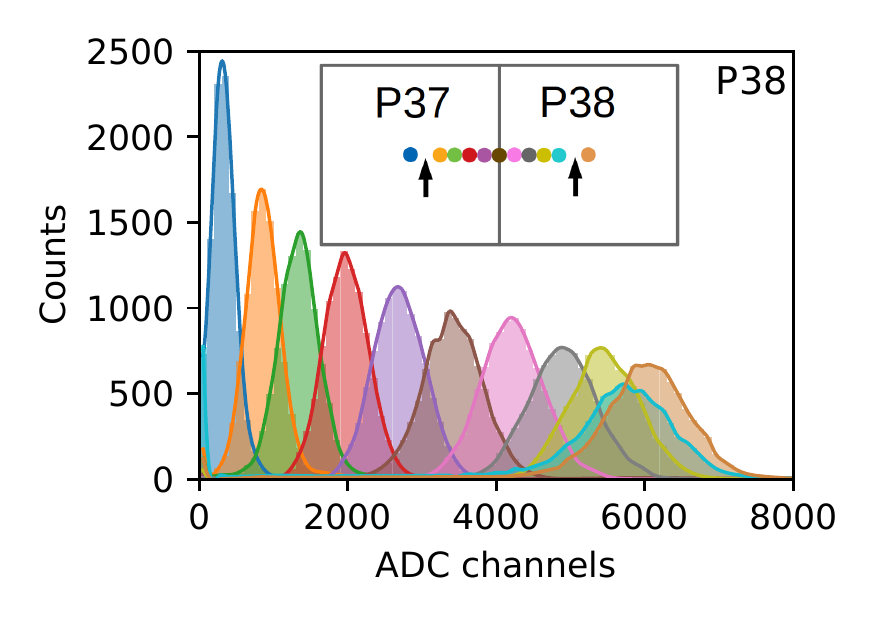}
			\caption{P38, collected charge}
			\label{subfig:prot_hor_P38}
		\end{subfigure}
		\begin{subfigure}[b]{0.47\textwidth}
			\includegraphics[width=1.\textwidth]{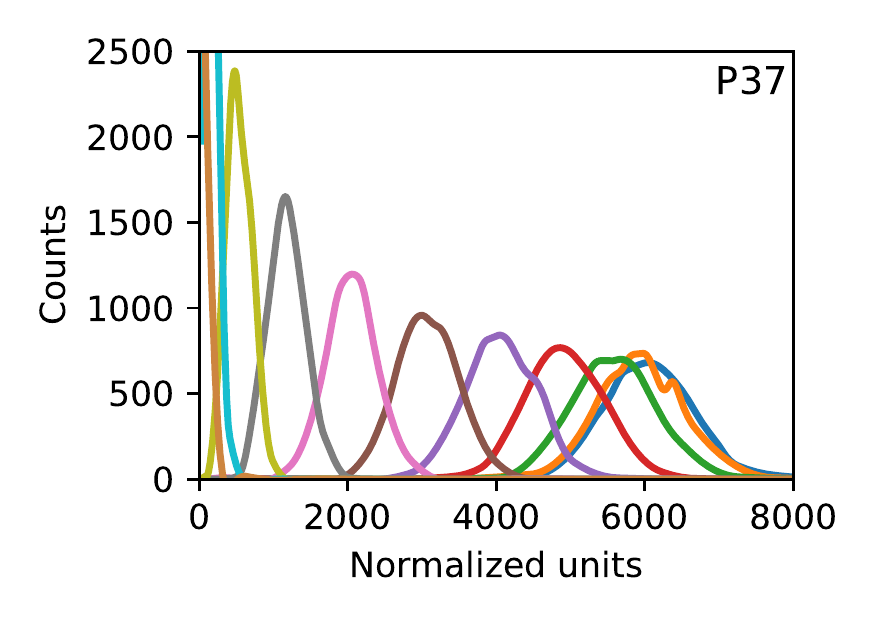}
			\caption{P37, \geant}
			\label{subfig:prot_hor_sim_P37}
		\end{subfigure}
		\begin{subfigure}[b]{0.47\textwidth}
			\includegraphics[width=1.\textwidth]{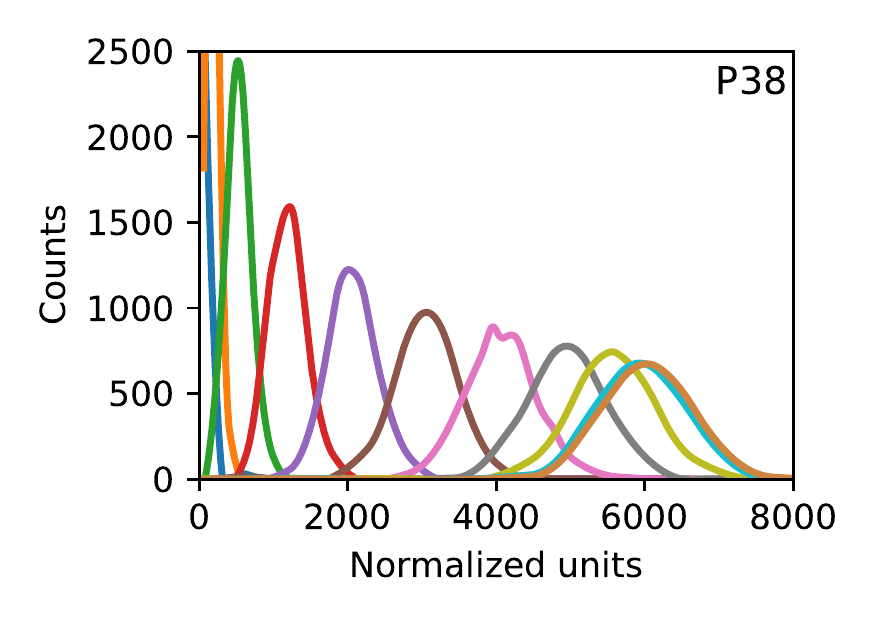}
			\caption{P38, \geant}
			\label{subfig:prot_hor_sim_P38}
		\end{subfigure}
		\caption{
			Horizontal scan, proton beam. The colors and beam 
			locations defined in the key (inset, top right 
			panel) apply to the spectra of gain-corrected 
			charge distributions (filled histograms, top 
			panels) and \geant~simulations of the 
			scintillation-light yield (open histograms, 
			bottom panels). Spectra taken at the arrowed 
			positions have been omitted for clarity, as they 
			substantially overlap the results from the 
			adjacent measurements. The normalized units were 
			chosen to match the simulated distributions to the 
			ADC spectra for the proton measurements at the pixel 
			centers.
			(For interpretation of the references to color 
			in this figure caption, the reader is referred 
			to the web version of this article.)
			}
		\label{figure:prot_hor}
        \end{center}
\end{figure}

Figure~\ref{figure:prot_hor} shows results from a horizontal scan of the SoNDe
module across the proton beam from the center of P37 to the center of P38 in
steps of 0.5\,mm. Also shown are \geant~simulations. For 11 scan positions, the
proton pulse-height spectra are displayed in Fig.~\ref{subfig:prot_hor_P37} 
(P37) and Fig.~\ref{subfig:prot_hor_P38} (P38) and the corresponding 
\geant-simulated scintillation-light yields are displayed in 
Fig.~\ref{subfig:prot_hor_sim_P37} (P37) and Fig.~\ref{subfig:prot_hor_sim_P38} 
(P38). The amount of scintillation light collected in a single pixel is clearly 
dependent upon the location of the proton beam. The amplitude of the signal is 
largest when scintillation light is produced at the center of the pixel and 
smallest when produced at the edge. Light produced at the boundary between two 
pixels is shared equally by both pixels. The simulations underestimate the 
amount of scintillation light spreading to the adjacent pixel closest to the 
particle interaction point by up to 15\% depending on the beam position. Note 
that the \geant~simulation does not include the $\sim$\SI{100}{\micro\meter} 
concavity of the MAPMT window but instead employs a constant air gap.

\begin{figure}[H]
	\begin{center}
		\begin{subfigure}[b]{0.47\textwidth}
			\includegraphics[width=1\textwidth]{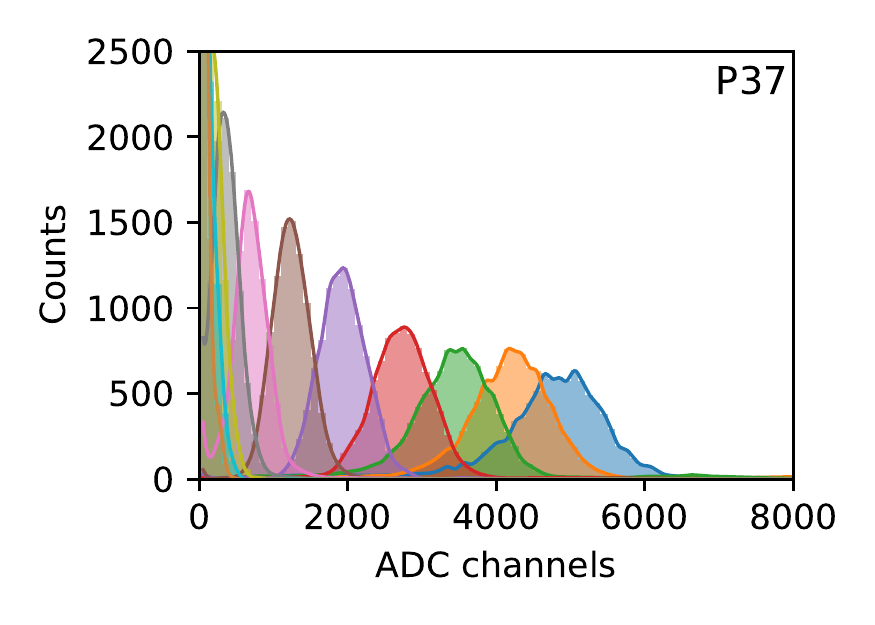}
			\caption{P37, collected charge}
			\label{subfig:deut_dia_P37}
		\end{subfigure}
		\begin{subfigure}[b]{0.47\textwidth}
			\includegraphics[width=1\textwidth]{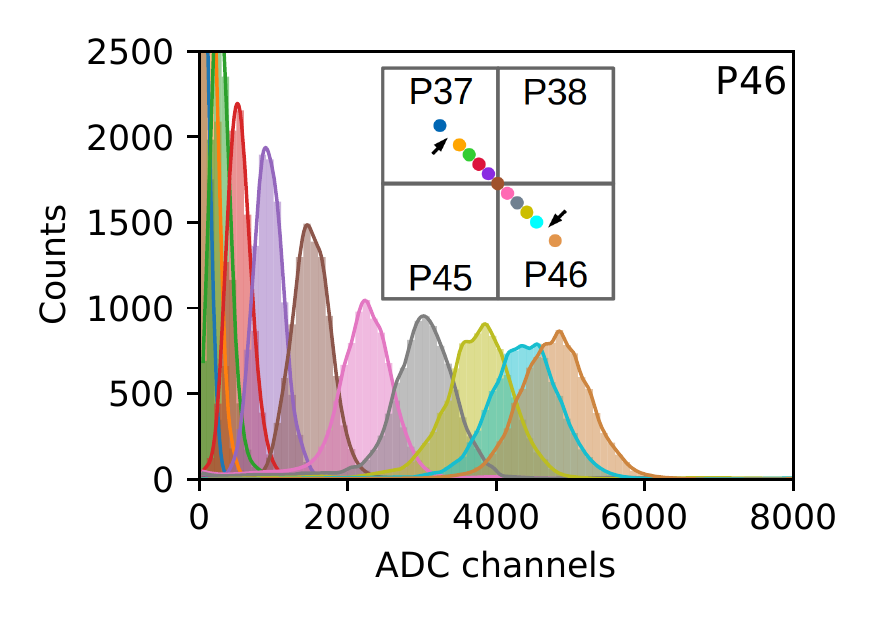}
			\caption{P46, collected charge}
			\label{subfig:deut_dia_P46}
		\end{subfigure}
		\begin{subfigure}[b]{0.47\textwidth}
			\includegraphics[width=1\textwidth]{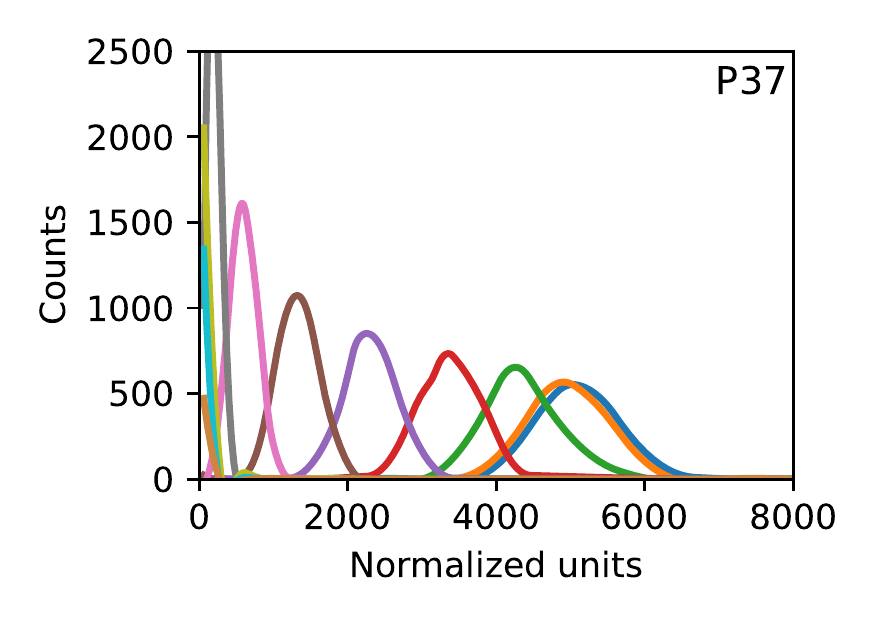}
			\caption{P37, \geant}
			\label{subfig:deut_dia_sim_P37}
		\end{subfigure}
		\begin{subfigure}[b]{0.47\textwidth}
			\includegraphics[width=1\textwidth]{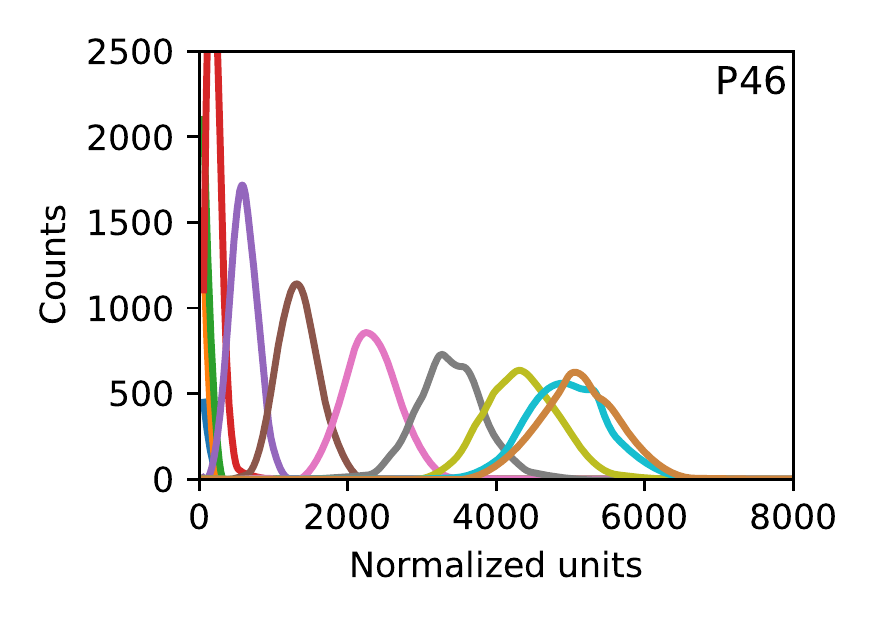}
			\caption{P46, \geant}
			\label{subfig:deut_dia_sim_P46}
		\end{subfigure}
		\caption{
			Diagonal scan, deuteron beam. The colors and beam 
			locations defined in the key (inset, top right 
			panel) apply to the spectra of gain-corrected 
			charge distributions (filled histograms, top 
			panels) and \geant~simulations of the 
			scintillation-light yield (open histograms, bottom 
			panels). Spectra taken at the arrowed positions 
			have been omitted for clarity, as they 
			substantially overlap the results from the adjacent 
			measurements. The normalized units were again 
			chosen to match the simulated distributions to the 
			ADC spectra for the proton measurements at the pixel 
			centers.
			(For interpretation of the references to color in this 
			figure caption, the reader is referred to the web 
			version of this article.)
			}
		\label{figure:deut_dia}
	\end{center}
\end{figure}

Figure~\ref{figure:deut_dia} shows results from a diagonal scan of the SoNDe
module across the 2.47\,MeV deuteron beam from the center of P37 to the center 
of P46 together with \geant~simulations. The scan was performed in a series of 
0.5\,mm horizontal and vertical steps, for an effective diagonal stepsize of 
0.71\,mm. For 13 scan positions, the deuteron pulse-height spectra are 
displayed in Fig.~\ref{subfig:deut_dia_P37} (P37) and 
Fig.~\ref{subfig:deut_dia_P46} (P46) and the corresponding \geant-simulated 
scintillation-light yields are displayed in Fig.~\ref{subfig:deut_dia_sim_P37} 
(P37) and Fig.~\ref{subfig:deut_dia_sim_P46} (P46). As anticipated, for a 
given pixel, the amplitude of the signal is largest when scintillation light 
is produced at the center of the pixel, and smallest when produced at the 
corner. Light produced at the corner of four pixels is shared equally by all 
four pixels. As before, the simulations underestimate the amount of 
scintillation light spreading to the pixels in close vicinity.

\begin{figure}[H]
        \begin{center}
		\begin{subfigure}[b]{0.55\textwidth}
			\includegraphics[width=1.1\textwidth]{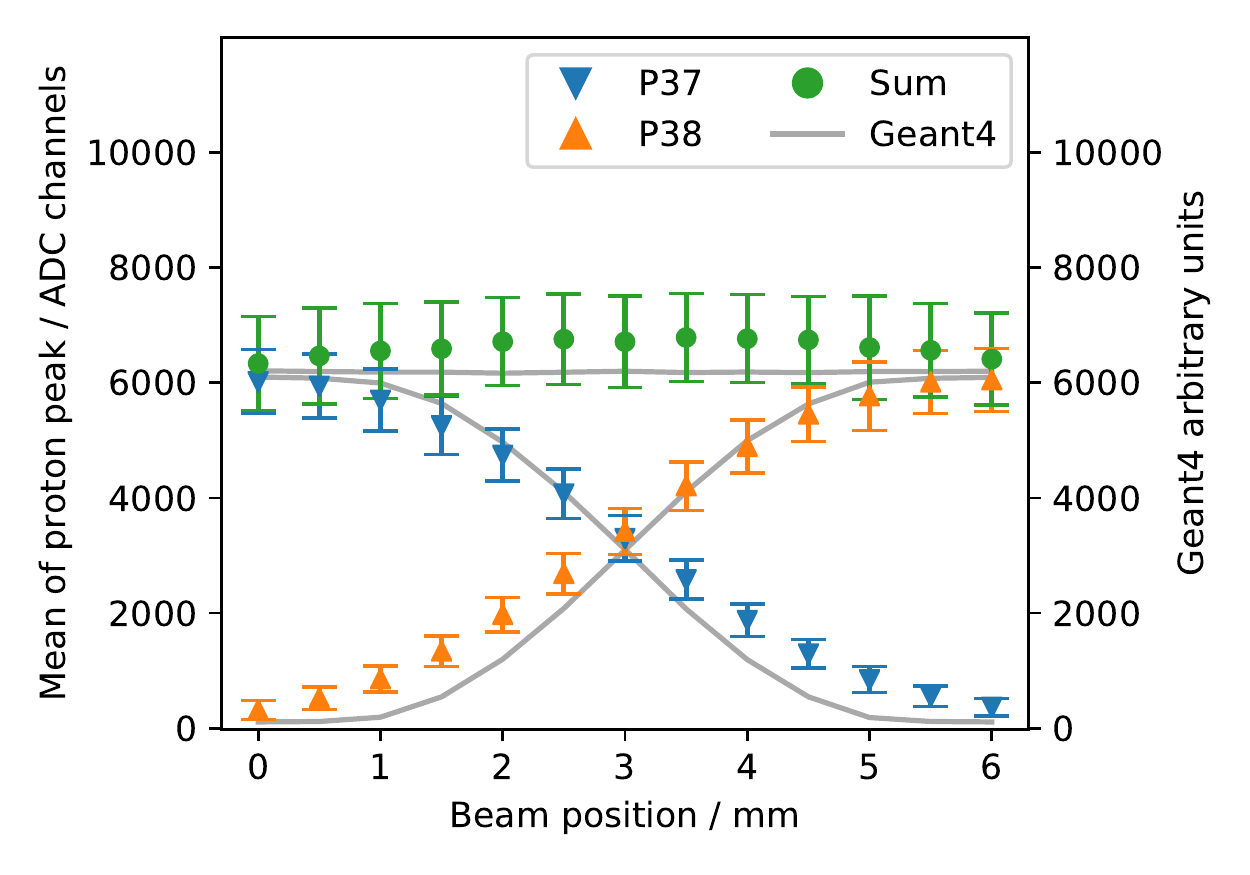}
			\caption{Proton beam}
			\label{subfig:lightshare_prot}
		\end{subfigure}
		\begin{subfigure}[b]{0.55\textwidth}
			\includegraphics[width=1.1\textwidth]{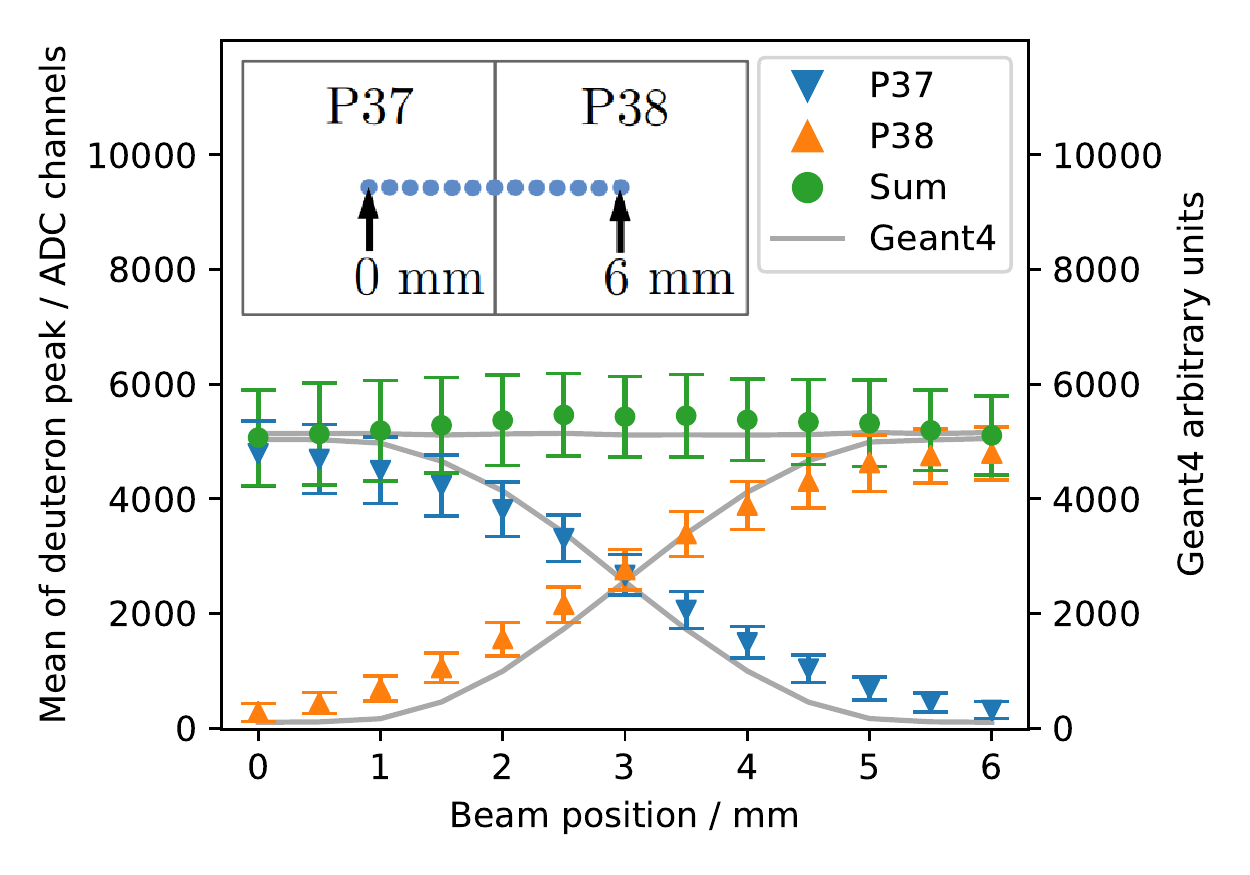}
			\caption{Deuteron beam}
			\label{subfig:lightshare_deut}
		\end{subfigure}
		\begin{subfigure}[b]{0.55\textwidth}
			\includegraphics[width=0.96\textwidth]{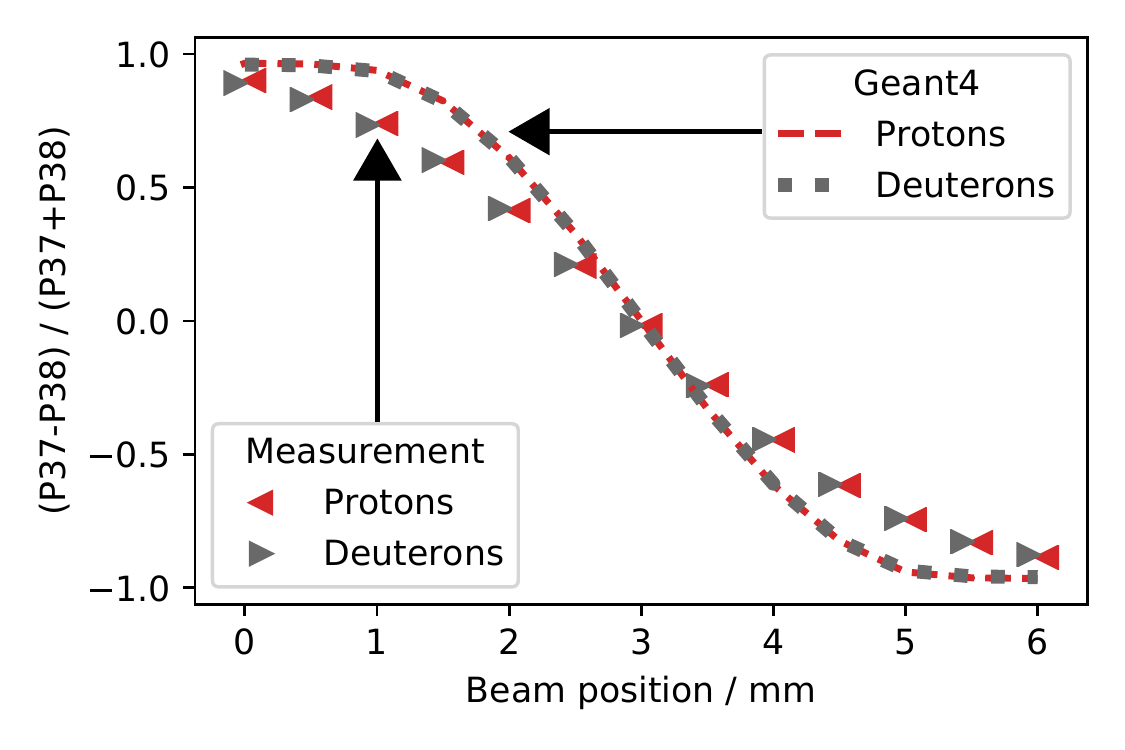}
			\caption{Normalized}
			\label{subfig:lightshare_normalized}
		\end{subfigure}
		\caption{
			Scintillation-light sharing. 
			\ref{subfig:lightshare_prot}: proton beam.
			\ref{subfig:lightshare_deut}: deuteron beam.
			Points are gain-corrected means of the charge 
			distributions corresponding to the division of 
			scintillation light between P37 and P38 for a 
			horizontal scan of the incident beam between pixel 
			centers in 0.5\,mm steps. The scan positions are 
			indicated in the inset to 
			\ref{subfig:lightshare_deut}. The uncertainties in 
			the means of the fitted Gaussian distributions are 
			smaller than the symbols. The error bars correspond 
			to $\pm 1\sigma$ of these fitted distributions. The 
			curves come from the corresponding 
			\geant~simulations of the scintillation light. The 
			simulations have been normalized to the measurements 
			as before.
			\ref{subfig:lightshare_normalized}: Light-sharing 
			ratios (P37$-$P38)/(P37+P38) derived from the plots 
			\ref{subfig:lightshare_prot} and 
			\ref{subfig:lightshare_deut}. The uncertainies are 
			smaller than the widths of the lines.
			}
		\label{figure:mean_yields_with_gain_correction}
        \end{center}
\end{figure}

Figure~\ref{figure:mean_yields_with_gain_correction} shows how the 
scintillation light was shared by adjacent pixels P37 and P38 as the SoNDe 
module was scanned horizontally across the proton 
(Fig.~\ref{subfig:lightshare_prot}) and deuteron beams 
(Fig.~\ref{subfig:lightshare_deut}).

\begin{figure}[H]
	\begin{center}
		\includegraphics[width=0.8\textwidth]{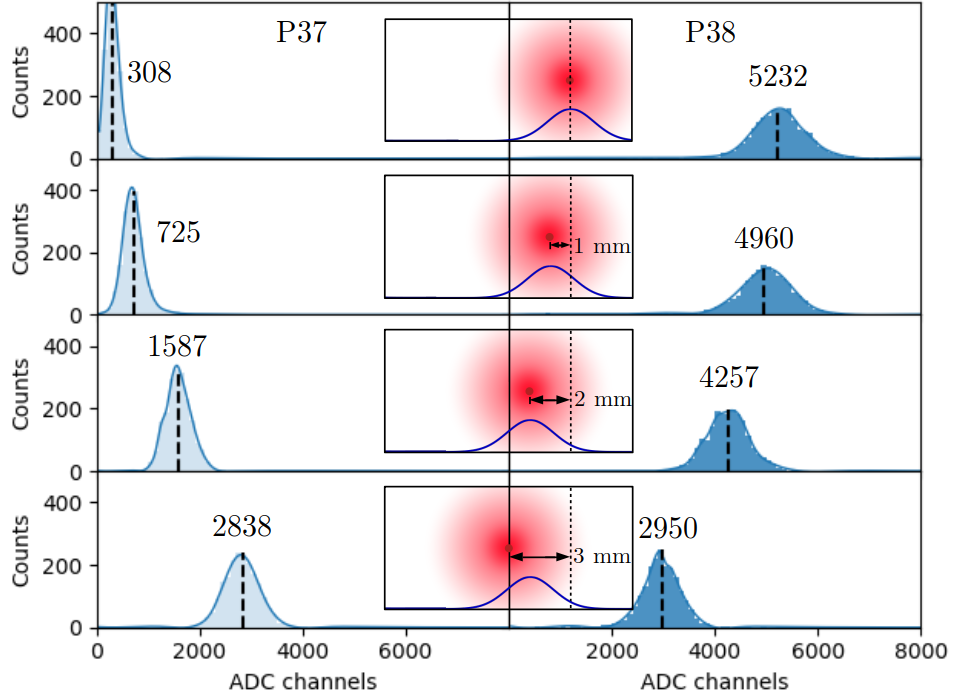}
		\caption{
			Division of signal across adjacent pixels, proton beam.
			Central insets: \geant-simulated scintillation-photon 
			distributions at the MAPMT photocathode (fuzzy red 
			circles) resulting from incident proton beams (dark red 
			dots) striking the SoNDe module at the three locations 
			shown lying on the line of the horizontal scan from P37 
			to P38. The Gaussian curves are the X-projections of 
			the simulated 2D distributions. The vertical line in 
			the center of each inset represents the boundary 
			between P37 and P38. Filled histograms: measured 
			gain-corrected charge distributions for P37 (lighter 
			blue, left column) and P38 (darker blue, right column).
			(For interpretation of the references to color in
			this figure caption, the reader is referred to the
			web version of this article.)
			}
		\label{figure:3_spectra}
	\end{center}
\end{figure}

\noindent	
Figures \ref{subfig:lightshare_prot} and \ref{subfig:lightshare_deut} show the 
means of the pulse-height distributions displayed as a function of beam 
position. The curves are spline fits to the corresponding \geant~simulations. 
The sum distributions show that the proton beam produced a factor of $\sim$1.25 
more scintillation light than the deuteron beam. The scan from P37 to P38 shows 
that light leakage to neighbouring pixels is relatively low close to pixel 
centers. Moving the particle beam from the center of P37 towards P38, $\sim$4\% 
of the total light yield is lost to P38 in the first mm. Across the boundary 
between the pixels, the light-loss gradient increases to 35\%/mm. Based upon the 
$\alpha$-particle scan results~\cite{rofors19}, it was anticipated that the sums 
of the gain-corrected charge distributions would be flat across the pixels and 
the boundary regions. Instead, the results have a slightly convex distribution 
centered at the pixel edge. This is because P37 and P38 together collect 
slightly more of the scintillation light produced from an event at the boundary 
between them than they collect from an event at the center of either pixel and 
the missing scintillation light is collected by the surrounding pixels
(Fig.~\ref{figure:3_spectra}). Figure \ref{subfig:lightshare_normalized} shows 
the light-sharing ratio between P37 and P38 (defined as the difference between 
the means of the signal distributions in the pixels divided by the sum) for 
both protons and deuterons. The overlap between the proton and deuteron data 
indicates that the light spreading mechanism is very similar for both particles.
The absolute difference between the data and the simulation is up to 20\%, 
greatest in the region between the center of a pixel and the edge. This 
difference could be due to scintillation-light spreading mechanisms which are not 
yet addressed in the simulation or even electronic crosstalk.

The standard mode of operation of SoNDe at ESS will be TOF mode in which every 
pixel exceeding its individual threshold will be time stamped, resulting in a 
data set of time-stamped pixel IDs without the underlying ADC information. Thus, 
knowledge of the behavior of adjacent pixels when the scintillation is registered 
in two or more of them is important. Figure~\ref{figure:3_spectra} shows the 
division of the signal in the SoNDe module as the proton beam was stepped across 
the boundary between adjacent pixels. The $\sim$\SI{100}{\micro\meter} diameter 
proton beam was simulated~\cite{annand20} using \geant to produce a distribution 
of scintillation light incident on the photocathode with a FWHM of $\sim$2\,mm. 
Given the 0.5\,mm mapping stepsize, this means that the majority of the 
scintillation light corresponding to irradiations at the center of a pixel or 
the first two horizontal steps towards a boundary is detected by the irradiated 
pixel. Due to the width of the photon distribution, an increasing amount of 
signal is registered by the adjacent pixel as the boundary is approached. In the 
top panel, the beam was centered on P38 resulting in a $\sim$(94/6) P38/P37 
division of the gain-corrected charge. In the second panel, the beam was 
translated 1\,mm towards the boundary between P38 and P37, resulting in a 
$\sim$(87/13) signal division. In the third panel, another 1\,mm shift closer to 
the P38/P37 boundary resulted in a $\sim$(73/27) division. In the bottom panel, 
the beam is incident on the boundary between pixels, resulting in a $\sim$(49/51)
signal division. It is thus possible that in the regions near the boundaries 
between pixels, a triggering event may result in a large amount of charge in 
adjacent pixels, especially near corners.

\begin{figure}[H]
	\centering
		\begin{subfigure}[b]{0.49\textwidth}
			\includegraphics[width=1\textwidth]{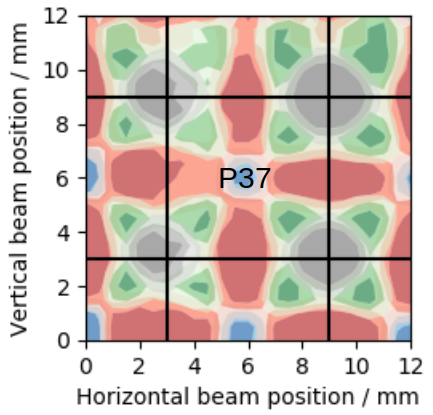}
			\caption{Spatial, threshold 600 ADC~channels}
			\label{subfig:600_channels}
		\end{subfigure}
		\begin{subfigure}[b]{0.50\textwidth}
			\includegraphics[width=1\textwidth]{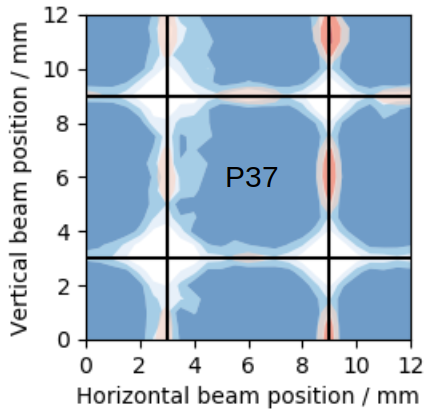}
			\caption{Spatial, threshold 2950 ADC~channels}
			\label{subfig:2950_channels}
		\end{subfigure}
		\begin{subfigure}[b]{0.65\textwidth}
			\includegraphics[width=1\textwidth]{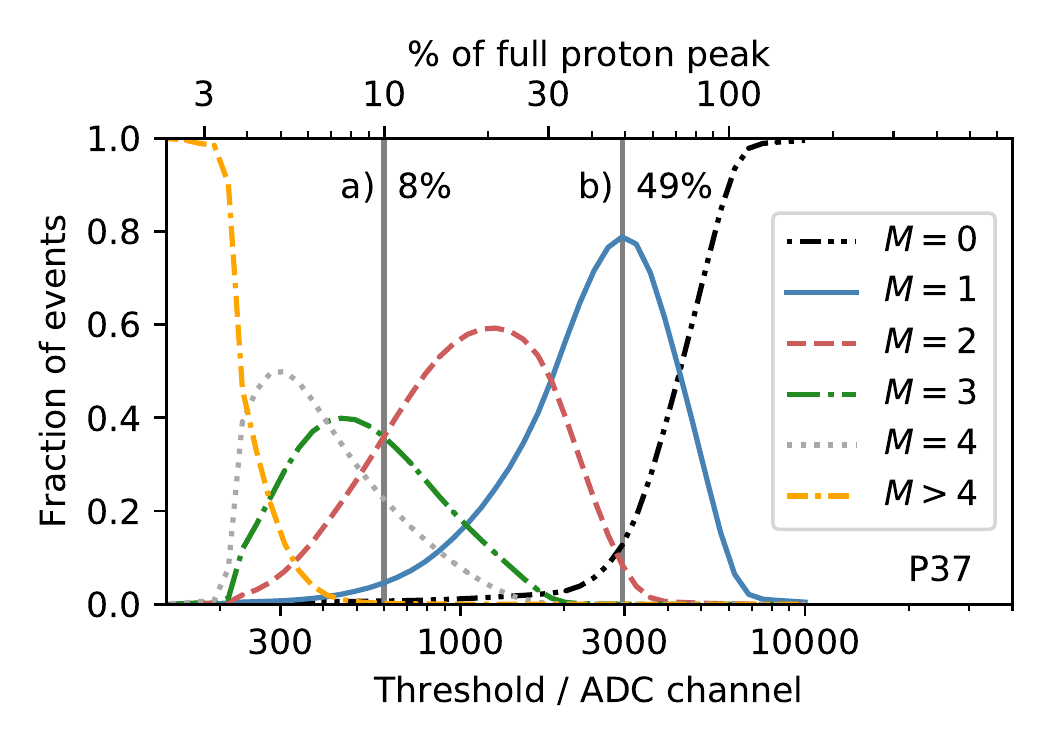}
			\vspace*{-8mm}
			\caption{Fractional}
			\label{subfig:mult_per_threshold}
		\end{subfigure}
		\begin{subfigure}[b]{0.25\textwidth}
			\hspace*{8mm}\includegraphics[width=\textwidth]{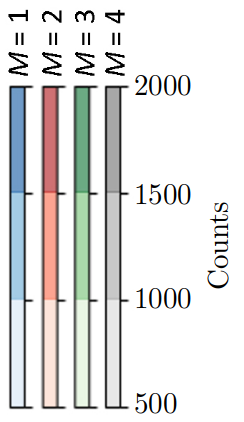}
			\vspace*{-2mm}
			\caption{Color key}
			\label{subfig:multkey}
		\end{subfigure}
		\caption{
			Multiplicity distributions for the proton beam 
			incident on P37 and the surrounding pixels. In 
			\ref{subfig:600_channels} (threshold 600~ADC 
			channels and \ref{subfig:2950_channels} 
			(threshold 2950~ADC channels, the black lines denote 
			the pixel boundaries. Blues indicate $M$~$=$~1 events, 
			reds indicate $M$~$=$~2 events, greens indicate 
			$M$~$=$~3 events, and greys indicate $M$~$=$~4 events. 
			The lighter the shade of the color, the fewer the 
			number of events. 
			\ref{subfig:mult_per_threshold} presents the fraction 
			of events registered in P37 for each multiplicity as 
			a function of threshold common to all pixels, with 
			the 600 and 2950~ADC channel thresholds shown as 
			vertical lines. 
			\ref{subfig:multkey} presents the color key for the 
			spatial distributions.
			(For interpretation of the references to color in this
			figure caption, the reader is referred to the web 
			version of this article.)
		}
		\label{figure:contour_maps_M_with_cuts}
\end{figure}

Knowledge of the behavior of the adjacent pixels as a function of threshold is 
important for the TOF mode of operation at ESS. In previous work~\cite{rofors19} 
determining responses to scans of $\sim$1\,mm FWHM beams of $\alpha$-particles, 
the hit multiplicity ($M$~$=$~1, $M$~$=$~2, etc.) for adjacent pixels as a 
function of the beam-spot position was measured. A hit was registered if a pixel 
amplitude exceeded a threshold which was variable. Here, the procedure was
repeated with the proton beam. A 26~$\times$~26 grid of proton-beam irradiations 
with a stepsize of 0.5\,mm in X and Y was performed. 
Figure~\ref{figure:contour_maps_M_with_cuts} displays results in the 
neighborhood of P37. Spatial distributions of multiplicity for software 
thresholds of 600 (Fig.~\ref{subfig:600_channels}) and 2950 
(Fig.~\ref{subfig:2950_channels}) channels are shown.  These thresholds 
correspond to 8\% and 49\% of the mean of the pixel-centered full-deposition 
proton peak, respectively. For a threshold of 600~channels, $M$~$=$~1 events are 
tightly constrained to within $\sim$1\,mm of pixel centers. Raising the threshold 
to 2950~channels results in the data being dominated by $M$~$=$~1 events to 
within $\sim$1\,mm of the pixel edges. The edges and corners are $M$~$=$~0 zones. 
The threshold clearly affects the multiplicity-dependent efficiency, and the 
2950~ADC channel threshold maximizes both the number of $M$~$=$~1 events detected 
and the area of the detector where the $M$~$=$~1 efficiency is high. 

Figure~\ref{subfig:mult_per_threshold} presents the threshold dependence of 
multiplicity for $M$~$=$~1$-$4. Each of the curves demonstrate clear maxima so 
that the relative contribution of a given $M$ can be maximized by suitable choice 
of threshold. For example, for the threshold of 2950~ADC channels optimized for 
$M$~$=$~1 events, $\sim$79\% of the events have $M$~$=$~1, $\sim$9\% have 
$M$~$=$~2, and a negligible number have $M$~$=$~3,4. The tradeoff is that 
$\sim$12\% of events have $M$~$=$~0, so that the consequence of operating the 
SoNDe module in $M$~$=$~1 mode is a loss of $\sim$12\% of events. A corresponding 
analysis of the deuteron data demonstrates the same behavior. The $M$~$=$~1 
optimal threshold cut for deuterons is ADC channel 2300 corresponding to 47\% of 
the mean of the pixel-centered full-deposition deuteron peak. Note that a 
threshold of at least 2500~ADC channels is necessary to completely discrimate 
against $\sim$1\,MeV $\gamma$-rays from a $^{60}$Co source, which is indicative 
of possible background contributions.

\section{Summary and Discussion}
\label{section:summary}

The position-dependent response of a SoNDe module, which consists of a 1\,mm 
thick sheet of GS20 scintillating glass coupled to a 64~pixel H12700A MAPMT 
has been measured using highly focused beams of protons and deuterons. The 
signal amplitudes from individual pixels were investigated as a function of 
beam position by stepping the module through the beams using a precision XY 
coordinate translator. The $\sim$\SI{100}{\micro\meter} diameter beams 
facilitated highly localized response mapping with a step size of 0.5\,mm. A 
detailed \geant~model of the SoNDe module greatly aided in the interpretation 
of these data and facilitated the calibration of the scintillation-light yield 
in GS20 as a function of beam energy for both beams 
(Fig.~\ref{figure:beam_details}). 

Spectra were gain corrected on a pixel-by-pixel basis using the data obtained 
when the beams were positioned at the center of each pixel. The signal 
amplitudes were highly dependent on the beam position 
(Figs.~\ref{figure:prot_hor} and \ref{figure:deut_dia}). The single-pixel 
signal was strongest when the beam was located at the pixel center. Moving 
the beam by $\sim$1\,mm from the center of a pixel towards a neighbouring 
pixel resulted in a $\sim$4\% leakage of the scintillation light to that 
pixel. As the pixel boundary was approached, the leakage gradient increased 
to $\sim$25\%/mm (Fig.~\ref{figure:mean_yields_with_gain_correction}). While 
the simulations underestimated the total amount of scintillation light shared 
across a pixel boundary, the overall agreement between the \geant~model and 
the data is very good. The amount of scintillation light produced in the GS20 
sheet by 2.48\,MeV protons was a factor of $\sim$1.25 greater than that 
produced by 2.47\,MeV deuterons. The spreading of light from protons and 
deuterons was indistinguishable. The \geant~simulation produced a 
visualization of the scintillation-photon distributions as a function of beam 
position (Fig.~\ref{figure:3_spectra}). The proton beam directed towards the 
central pixel region resulted in little signal in an adjacent pixel. However, 
within $\sim$1\,mm of the boundary, at least 40\% of the scintillation light 
was registered in the adjacent pixel. When in use at ESS, SoNDe will record 
only the time-stamped pixel IDs for every pixel exceeding its individual 
threshold. Thus, in the pixel-boundary region, double counting can occur. The 
effect of the pixel threshold on double and even higher-order counting (the 
hit multiplicity) was studied as a function of beam position and threshold
(Fig.~\ref{figure:contour_maps_M_with_cuts}). At a threshold of $\sim$50\% of 
the mean of the proton full-deposition peak, $\sim$80\% of the beam protons 
were registered in a single pixel. Of the remaining protons, $\sim$10\% were 
double counted and $\sim$10\% were not detected. The double-counted protons 
were confined to regions within $\sim$0.5\,mm of pixel edges, while 
undetected protons were confined to regions within $\sim$1\,mm of pixel 
corners. Thus, operated in this mode, the active area of the SoNDe module 
(87\% of the MAPMT surface) provides a position resolution of $\sim$5\,mm in 
X and Y and a detection efficiency of $\sim$80\% for 2.49\,MeV protons. 
Increasing the threshold further simply resulted in a further reduction in 
the single-pixel event-detection efficiency and sensitive area. 

\section*{Acknowledgements}
\label{acknowledgements}

The support of the European Union via the Horizon 2020 Solid-State Neutron 
Detector Project, Proposal ID 654124, and the BrightnESS Project, Proposal 
ID 676548 is acknowledged. The support of the UK Science and Technology 
Facilities Council (Grant No. ST/P004458/1) and UK Engineering and Physical 
Sciences Research Council Centre for Doctoral Training in Intelligent Sensing 
and Measurement (Grant No. EP/L016753/1) are also acknowledged, as is 
Strategic Accelerator Support from the Engineering Faculty (LTH) of Lund 
University (Grant No. STYR 2019/1508).

\newpage

\bibliographystyle{elsarticle-num}

\end{document}